\documentclass[10pt,onecolumn,aps,prd,preprintnumbers,showpacs,superscriptaddress,nofootinbib,amsmath,amssymb,floats,floatfix,showkeys,notitlepage,longbibliography]{revtex4-2}
\usepackage{comment}
\usepackage{lipsum}
\usepackage{graphicx}
\usepackage{subfigure}
\usepackage{palatino}
\usepackage{sans}
\usepackage{array}
\usepackage[toc,page]{appendix}
\usepackage[normalem]{ulem}
\usepackage{adjustbox}
\usepackage{latexsym}
\usepackage{amsmath}
\usepackage{amssymb}
\usepackage{amsfonts}
\numberwithin{equation}{section}
\usepackage{mathrsfs}
\usepackage{physics}
\usepackage{dcolumn}
\usepackage{bm}
\usepackage{tikz}
\usetikzlibrary{decorations.pathmorphing}
\usepackage{pgfplots}
\pgfplotsset{compat=1.18}
\usepackage{bigints}
\usepackage{array,tabularx,multirow,booktabs}
\usepackage[tracking=true]{microtype}
\usepackage{soul} 
\SetTracking{}{500}
\SetTracking{encoding={*}, shape=sc}{40}
\UseRawInputEncoding 
\allowdisplaybreaks
\usepackage[utf8]{inputenc}
\usepackage{xcolor} 
\usepackage{babel}
\usepackage{etoolbox}
\usepackage{hyperref}
\hypersetup{colorlinks=true,linkcolor=blue,urlcolor=blue,citecolor=blue}
\usepackage{orcidlink}

\begin{document} \sloppy

\title{Optical-area minimum method for static spherical black hole shadows}

\author{Vitalii Vertogradov
\orcidlink{0000-0002-5096-7696}}
\email{vdvertogradov@gmail.com}
\affiliation{Wilczek Quantum Center, Shanghai Institute for Advanced Studies, Shanghai, 201315, China.}
\affiliation{University of Science and Technology of China, Hefei, 230026, China}

\author{Nikko John Leo S. Lobos \orcidlink{0000-0001-6976-8462}}
\email{nslobos@ust.edu.ph}
\affiliation{Electronics Engineering Department, University of Santo Tomas, Espa\~na Boulevard, Sampaloc, Manila 1008, Philippines}

\author{Ali \"Ovg\"un
\orcidlink{0000-0002-9889-342X}
}
\email{ali.ovgun@emu.edu.tr}
\affiliation{Physics Department, Eastern Mediterranean
University, Famagusta, 99628 North Cyprus, via Mersin 10, Turkiye.}

\author{Reggie C. Pantig \orcidlink{0000-0002-3101-8591}}
\email{rcpantig@mapua.edu.ph}
\affiliation{Physics Department, School of Foundational Studies and Education,
Map\'ua University, 658 Muralla St., Intramuros, Manila 1002, Philippines}

\date{\today}

\begin{abstract}
We formulate a global optical-area method for shadows of static, spherically
symmetric black holes. For the metric
\(ds^{2}=-A(r)dt^{2}+B(r)dr^{2}+C(r)d\Omega^{2}\), spherical sections of the
optical geometry have area
\(\mathcal{A}_{\rm opt}=4\pi C/A\). A null ray with impact parameter \(b\)
can cross a spherical section only if \(b^{2}\leq C/A\). The capture threshold
is fixed by the infimum of \(C/A\) on the connected interval between the
observer and the black hole horizon. When attained at an interior point, the
infimum gives \(b_{\rm sh}^{2}=\min(C/A)\), while a static observer outside
the controlling minimum, on the inward-sky branch, measures
\(\sin^{2}\alpha_{\rm sh}=\mathcal{A}_{*}/
\mathcal{A}_{\rm opt}(r_{\rm o})\). The usual photon-sphere equation follows
when the minimum occurs at a smooth interior point. Exponential instability
additionally requires the minimum to be nondegenerate. The local
optical-radius and photon-sphere formulas are established results. Our
contribution is to organize them into an observer-to-horizon global selection
rule that compares all stationary candidates and relevant endpoint limits.
The radial function
\(B(r)\) does not affect the shadow angle, although it enters the
coordinate-time instability rate, whose numerical value also depends on the
normalization of the static time coordinate. We derive compact first- and
second-order formulas for deformed metrics, demonstrate candidate comparison
with a synthetic two-minimum profile, and apply the construction to
Reissner--Nordstr\"om, Bardeen, charged dilaton, and Kottler black holes. An
explicit transformation of the charged-dilaton
example from a nonareal to an areal radial coordinate verifies
radial-coordinate invariance, while the Kottler example probes a
nonasymptotically flat static region.
\end{abstract}


\maketitle

\section{Introduction}
\label{sec:introduction}

The boundary of a black hole shadow is generated by null rays that separate
capture from escape and is naturally described in terms of an escape cone or
a critical impact parameter
\cite{Synge:1966okc,Virbhadra:1999nm,Bozza:2002zj,Bozza:2010xqn,
Perlick:2021aok}. In a static
and spherically symmetric spacetime, the
standard calculation introduces the conserved photon energy and angular
momentum, constructs a radial effective potential, and solves the conditions
for an unstable circular null orbit or photon surface
\cite{Claudel:2000yi,Perlick:2021aok}. This procedure is efficient when the
metric has a single exterior photon sphere, but its local stationarity
condition is less informative when several circular null orbits are present.
Every stationary radius is then a candidate, and a separate global argument
is needed to identify the one that controls capture from a specified
observer
\cite{Cunha:2017qtt,Cunha:2017eoe,Shaikh:2019itn,
Gan:2021pwu,Guo:2022muy,Guo:2022ghl}.

Horizon-scale observations have made the relation between geometric shadows
and measured ring diameters phenomenologically important. The Event Horizon
Telescope images of M87* and Sagittarius~A* provide the principal current
baselines
\cite{EventHorizonTelescope:2019dse,EventHorizonTelescope:2019ggy,
EventHorizonTelescope:2022wkp,EventHorizonTelescope:2022xqj}. Their use in tests
of gravity requires care because the capture shadow, photon ring, lensing
rings, and emission-dependent brightness features are distinct, and different
metrics can produce similar shadow sizes
\cite{Gralla:2019xty,Gralla:2020srx,Lima:2021las}. Shadow-size
constraints and theory surveys illustrate both the promise and the model
dependence of such tests
\cite{EventHorizonTelescope:2020qrl,EventHorizonTelescope:2021dqv,
Vagnozzi:2022moj}.

Optical geometry has long provided a geometric description of null rays in
static spacetimes. In spherical symmetry, the optical areal radius and its
stationary points have been used to characterize circular photon orbits,
their stability, and shadow radii
\cite{Hasse:2001by,Cvetic:2016bxi,Qiao:2022jlu,Qiao:2022hfv,Baines:2023bgm}. Finite-distance
shadow angles for static spherical geometries are likewise established in
the literature \cite{Perlick:2021aok}. Thus neither the ratio \(C/A\) nor
the local photon-sphere condition is introduced here as a new invariant.
Related optical-geometric methods have also been developed for weak and
strong lensing through Gauss--Bonnet and Randers constructions
\cite{Gibbons:2008rj,Werner:2012rc}.

Here we reorganize the calculation around the spherical areas of the optical
geometry. We call the resulting prescription the optical-area minimum method
(OAMM). Its primary quantity is
\[
\mathcal{S}(r)=\frac{C(r)}{A(r)},
\]
which is the squared areal radius of a spherical section in the optical
metric. A photon with impact parameter \(b\) can traverse such a section only
when \(b^{2}\leq\mathcal{S}(r)\). A ray sent inward from an observer must pass
all sections between the observer and the horizon, so the smallest accessible
value of \(\mathcal{S}\) fixes the capture threshold. The familiar
photon-sphere equation is recovered when this global minimum is attained at a
smooth interior point, but it is not used as the starting assumption.

The emphasis of OAMM is the observer-to-horizon global selection rule. All
admissible stationary values and relevant endpoint limits are compared on the
connected static interval accessible to the observer. This formulation makes
the shell-crossing argument, the distinction between a minimum and an
endpoint infimum, and the finite-distance optical-area ratio explicit in one
scalar function. The construction is algebraically equivalent to the
standard null-geodesic and effective-potential treatments. It does not
introduce a new local photon-sphere equation, a new critical impact parameter,
or a new finite-distance tetrad relation. In particular, optical-radius
extrema and their photon-orbit interpretation are established in
Ref. \cite{Hasse:2001by}, the arbitrary-distance static-observer relation
is reviewed in Ref. \cite{Perlick:2021aok}, multiple exterior photon
spheres and their competing lensing barriers are analyzed in
Refs. \cite{Guo:2022muy,Guo:2022ghl}, and systematic perturbative shadow
expansions are developed in
Refs. \cite{Kobialko:2024zhc,Pantig:2025deu}. The distinct claim made here
is therefore limited to using the observer-accessible infimum, including
relevant endpoint limits, as the explicit organizing algorithm and combining
that selection rule with the finite-distance area ratio and extremal-value
expansion in one scalar framework. A conventional effective-potential
calculation gives the same prediction when all barriers and accessibility
conditions are analyzed globally.

The broader shadow literature includes rotating and non-Kerr compact objects,
regular black holes, and phenomenological deformations
\cite{Bambi:2008jg,Hioki:2009na,Amarilla:2010zq,
Amarilla:2011fx,Grenzebach:2014fha,Abdujabbarov:2016hnw,
Amir:2016cen,Tsukamoto:2017fxq,Johannsen:2010ru}. Those settings generally
require a direction-dependent shadow curve rather than the single scalar
minimum used here, but they motivate a precise statement of the symmetry and
observer assumptions behind OAMM. Recent model applications in, for example,
Annals of Physics likewise use shadow observables together with lensing or
greybody information to test deformed black hole metrics
\cite{Pantig:2021zqe,Pantig:2022ely}.

This viewpoint has four useful consequences. First, it supplies a direct
selection rule when several stationary light radii occur. Second, it gives
the finite-distance shadow angle as a ratio of optical areas. Third, it makes
clear that \(A(r)\) and \(C(r)\), but not \(B(r)\), determine the shadow
silhouette seen by a static observer. Fourth, it yields particularly simple
perturbative formulas. The first-order change in the shadow radius can be
computed without first finding the displacement of the critical radius.

Systematic first- and second-order perturbation theories for shadow and
confining-surface observables have recently been developed for static
spherical metrics \cite{Kobialko:2024zhc,Pantig:2025deu}. The formulas
below isolate the vacuum-null specialization as an expansion of the extremal
value of the single scalar \(\mathcal{S}=C/A\). Their practical simplification
is that the first-order change of the minimum value is evaluated at the
undeformed critical radius, while the displacement of that radius first
enters the minimum value at second order.

We develop the formalism for a general static spherical metric, relate the
global minimum to the unstable circular null orbit and its instability rate,
derive finite-distance celestial observables, and obtain perturbative
expressions through second order. We first illustrate genuine competition
between two admissible minima with a synthetic asymptotically flat profile.
We then implement the method for the
Reissner--Nordstr\"om, Bardeen, charged dilaton, and Kottler geometries. The
last two applications respectively test a nonareal radial coordinate and a
nonasymptotically flat static region.

Throughout, we use geometrized units \(G=c=1\).

\section{Optical geometry and the global area minimum}
\label{sec:optical-geometry}

Consider a static and spherically symmetric line element
\begin{equation}
ds^{2}
=-A(r)dt^{2}
+B(r)dr^{2}
+C(r)\left(d\theta^{2}+\sin^{2}\theta\,d\phi^{2}\right),
\label{eq_metric}
\end{equation}
where \(A>0\), \(B>0\), and \(C>0\) in the static region of interest. The
observer is located at \(r=r_{\rm o}\), and the black hole horizon is at
\(r=r_{\rm h}<r_{\rm o}\). Unless stated otherwise, we assume that the
observer lies outside the optical-area minimum that controls capture.

For a null curve, Eq. \eqref{eq_metric} gives
\begin{equation}
dt^{2}
=\frac{B(r)}{A(r)}dr^{2}
+\frac{C(r)}{A(r)}d\Omega^{2},
\qquad
d\Omega^{2}=d\theta^{2}+\sin^{2}\theta\,d\phi^{2}.
\label{eq_null-condition}
\end{equation}
The corresponding spatial optical metric is therefore
\begin{equation}
d\ell_{\rm opt}^{2}
=\frac{B(r)}{A(r)}dr^{2}
+\frac{C(r)}{A(r)}d\Omega^{2}.
\label{eq_optical-metric}
\end{equation}
This conformal spatial metric is the standard optical geometry of a static
spacetime \cite{Hasse:2001by,Perlick:2004tq,Qiao:2022hfv}.
A spherical section at fixed \(r\) has optical area
\begin{equation}
\mathcal{A}_{\rm opt}(r)
=4\pi\frac{C(r)}{A(r)}.
\label{eq_optical-area}
\end{equation}
It is convenient to define the optical areal radius and its square by
\begin{equation}
\mathcal{R}(r)
=\sqrt{\frac{\mathcal{A}_{\rm opt}(r)}{4\pi}}
=\sqrt{\frac{C(r)}{A(r)}},
\qquad
\mathcal{S}(r)=\mathcal{R}^{2}(r)=\frac{C(r)}{A(r)}.
\label{eq_R-and-S}
\end{equation}

The central construction is the smallest value of \(\mathcal{S}\) on the
connected static path from the observer toward the black hole horizon. It is
defined by
\begin{equation}
\mathcal{S}_{*}
=\inf_{r_{\rm h}<r<r_{\rm o}}\mathcal{S}(r),
\qquad
\mathcal{A}_{*}=4\pi\mathcal{S}_{*}.
\label{eq_global-infimum}
\end{equation}
The use of the infimum keeps the statement valid when the limiting value is
approached at an endpoint. approached at an endpoint. An explicit singular-horizon realization of an
endpoint-controlled shadow is analyzed in Ref. \cite{PANTIG2026140643}. In the
black hole exteriors considered below, the relevant value is attained at an
interior radius \(r=r_{*}\), and \(\mathcal{S}_{*}\) is an ordinary global
minimum.

For a nonextremal asymptotically flat black hole with finite, nonzero
\(C(r_{\rm h})\), one commonly has
\begin{equation}
\lim_{r\to r_{\rm h}^{+}}\mathcal{R}(r)=+\infty,
\qquad
\lim_{r\to\infty}\mathcal{R}(r)=+\infty.
\label{eq_asymptotic-limits}
\end{equation}
A minimum must then occur between the horizon and infinity. A similar
two-sided divergence occurs between the black hole and cosmological horizons
of the subextremal Kottler spacetime.

If the global minimum is attained at a smooth interior point, the general
necessary conditions are
\begin{equation}
\mathcal{S}'(r_{*})=0,
\qquad
\mathcal{S}''(r_{*})\geq 0.
\label{eq_min-conditions}
\end{equation}
For an ordinary unstable circular null orbit and the quadratic instability
analysis in Sec. \ref{sec:capture}, we make the additional nondegeneracy
assumption \(\mathcal{S}''(r_{*})>0\). A degenerate minimum with
\(\mathcal{S}''(r_{*})=0\) requires expansion to the first nonzero higher
derivative and need not produce exponential instability.
Using \(\mathcal{S}=C/A\), the stationarity condition becomes
\begin{equation}
A(r_{*})C'(r_{*})-C(r_{*})A'(r_{*})=0.
\label{eq_stationarity}
\end{equation}
Equation \eqref{eq_stationarity} supplies candidates only. When more than one
admissible root exists, their values of \(C/A\) must be compared over the full
observer-to-horizon interval. This global comparison is an essential part of
the method.

The construction is invariant under any smooth monotonic change of radial
coordinate. Such a relabeling changes the functional forms assigned to
\(A\), \(B\), and \(C\), but it does not change the values of \(C/A\) on the
physical spherical sections or their ordering along the radial path.

\section{Capture threshold and critical null orbit}
\label{sec:capture}

Spherical symmetry allows every photon trajectory to be placed in the
equatorial plane, \(\theta=\pi/2\). Because coordinate time is optical arc
length along a null ray, Eq. \eqref{eq_optical-metric} yields
\begin{equation}
1
=\frac{B}{A}\left(\frac{dr}{dt}\right)^{2}
+\mathcal{R}^{2}(r)\left(\frac{d\phi}{dt}\right)^{2}.
\label{eq_optical-motion}
\end{equation}
Rotational symmetry of the optical metric gives the conserved quantity
\begin{equation}
j=\mathcal{R}^{2}(r)\frac{d\phi}{dt}.
\label{eq_optical-angular-momentum}
\end{equation}

Let \(\lambda\) be an affine parameter. Staticity and spherical symmetry give
\begin{equation}
E=A(r)\frac{dt}{d\lambda},
\qquad
L=C(r)\frac{d\phi}{d\lambda},
\qquad
b=\frac{L}{E}.
\label{eq_constants}
\end{equation}
These are the standard first integrals for null geodesics in a static
spherical metric \cite{Perlick:2021aok}.
It follows directly that
\begin{equation}
\mathcal{R}^{2}\frac{d\phi}{dt}
=\frac{C}{A}
\frac{d\phi/d\lambda}{dt/d\lambda}
=\frac{L}{E}
=b,
\label{eq_j-equals-b}
\end{equation}
so the conserved optical angular momentum is the usual impact parameter.
Substitution into Eq. \eqref{eq_optical-motion} gives
\begin{equation}
\frac{B}{A}\left(\frac{dr}{dt}\right)^{2}
=1-\frac{b^{2}}{\mathcal{S}(r)}.
\label{eq_radial-crossing}
\end{equation}
This equation is used here as a shell-crossing condition rather than as an
effective-potential construction. A ray can cross a spherical section only
if
\begin{equation}
b^{2}\leq\mathcal{S}(r).
\label{eq_crossing-condition}
\end{equation}
An inward ray reaches the horizon only if this inequality holds at every
intermediate radius. The capture threshold is therefore
\begin{equation}
b_{\rm sh}^{2}
=\mathcal{S}_{*}
=\inf_{r_{\rm h}<r<r_{\rm o}}\frac{C(r)}{A(r)}
=\frac{\mathcal{A}_{*}}{4\pi}
.
\label{eq_central-result}
\end{equation}
In the applications below, the observer lies outside an attained controlling
minimum \(r_{*}<r_{\rm o}\), so its value is independent of the observer
radius. At the observer, Eq. \eqref{eq_radial-crossing} first imposes the
local admissibility condition
\begin{equation}
0\leq b^{2}\leq \mathcal{S}(r_{\rm o})
\equiv\mathcal{S}_{\rm o}.
\label{eq_observer-admissibility}
\end{equation}
When the infimum is attained at an interior nondegenerate minimum, the
locally admissible rays divide into three impact-parameter classes.
\begin{equation}
\begin{array}{lll}
0\leq b^{2}<\mathcal{S}_{*} &\Longrightarrow& \text{capture by the black hole},\\
b^{2}=\mathcal{S}_{*} &\Longrightarrow& \text{critical separatrix},\\
\mathcal{S}_{*}<b^{2}\leq\mathcal{S}_{\rm o}
&\Longrightarrow& \text{turning point before the horizon}.
\end{array}
\label{eq_ray-classification}
\end{equation}
For \(b^{2}>\mathcal{S}_{\rm o}\), the radial equation is already negative at
the observer, so no real inward null direction with that impact parameter can
be launched there.
The equality case approaches the critical circular orbit asymptotically and
should not be classified as a horizon-crossing captured ray. If the infimum
is approached only at an endpoint, there is no interior circular separatrix.
The equality case must instead be determined from the local endpoint
geometry. The two strict inequalities retain their shell-crossing meaning
within the observer admissibility range \eqref{eq_observer-admissibility}.

The equality case explains the connection with the usual unstable circular
null orbit. Introduce optical proper radial distance
\begin{equation}
d\sigma=\sqrt{\frac{B(r)}{A(r)}}\,dr.
\label{eq_sigma}
\end{equation}
The equatorial optical metric and radial equation become
\begin{equation}
d\ell_{\rm opt}^{2}=d\sigma^{2}+\mathcal{R}^{2}(\sigma)d\phi^{2},
\qquad
\left(\frac{d\sigma}{dt}\right)^{2}
=1-\frac{b^{2}}{\mathcal{R}^{2}(\sigma)}.
\label{eq_sigma-motion}
\end{equation}
At the nondegenerate minimum assumed for the remainder of this subsection,
\begin{equation}
\left.\frac{d\mathcal{R}}{d\sigma}\right|_{*}=0,
\qquad
\left.\frac{d^{2}\mathcal{R}}{d\sigma^{2}}\right|_{*}>0.
\label{eq_R-sigma-minimum}
\end{equation}
Writing \(\delta\sigma=\sigma-\sigma_{*}\) and setting
\(b=\mathcal{R}_{*}\), a local expansion gives
\begin{equation}
\left(\frac{d\delta\sigma}{dt}\right)^{2}
=\frac{\mathcal{R}_{*}''}{\mathcal{R}_{*}}\,
\delta\sigma^{2}
+\mathcal{O}(\delta\sigma^{3}),
\label{eq_local-instability}
\end{equation}
where the double prime in this equation denotes differentiation with respect
to \(\sigma\). Thus
\begin{equation}
\delta\sigma(t)
\simeq
\delta\sigma_{0}
\exp\!\left[\pm\lambda_{*}(t-t_{0})\right],
\qquad
\lambda_{*}^{2}
=\frac{\mathcal{R}_{*}''}{\mathcal{R}_{*}}.
\label{eq_lambda-def}
\end{equation}
The critical ray approaches the minimum asymptotically in one time direction
and departs exponentially in the other. The unstable circular null orbit
therefore follows from a nondegenerate interior minimum. For a degenerate
minimum, the quadratic coefficient in Eq. \eqref{eq_local-instability}
vanishes and the leading higher-order equation generally gives nonexponential
critical behavior.

For calculations in the original radial coordinate,
\begin{equation}
\lambda_{*}^{2}
=
\left.
\frac{A}{B}
\frac{\mathcal{S}''}{2\mathcal{S}}
\right|_{r=r_{*}}
.
\label{eq_lambda-coordinate}
\end{equation}
This coordinate-time Lyapunov exponent is the usual instability test
for circular null geodesics \cite{Cardoso:2008bp,Decanini:2010fz}. Unlike
the shadow radius, the rate involves \(B(r)\) as well as \(A(r)\), \(C(r)\),
and their derivatives. It also depends on the normalization of the static
time coordinate. Under \(\bar t=\kappa t\) with constant \(\kappa>0\), one
has
\begin{equation}
\bar A=\frac{A}{\kappa^{2}},
\qquad
\bar b_{\rm sh}=\kappa b_{\rm sh},
\qquad
\bar\lambda_{*}=\frac{\lambda_{*}}{\kappa}.
\label{eq_time-rescaling}
\end{equation}
Comparisons of \(\lambda_{*}\) between spacetimes therefore require a stated
time-normalization convention. Alternatively, a static observer at
\(r_{\rm o}\) may quote the rate per unit proper time,
\(\lambda_{*}^{({\rm o})}=\lambda_{*}/\sqrt{A_{\rm o}}\), which is invariant
under the constant rescaling above. The pair
\((b_{\rm sh},\lambda_{*})\), with a specified time normalization, separates
the location of the shadow edge from the coordinate-time rate at which
nearby critical rays leave it.

The same instability scale also appears in eikonal connections among null
orbits, strong-deflection lensing, and quasinormal modes
\cite{Stefanov:2010xz,Yang:2012he}. Extensions of shadow--quasinormal-mode
relations to rotating geometries require additional orbital frequencies and
are correspondingly less reducible to a single spherical optical radius
\cite{Yang:2021zqy}.

\section{Finite-distance shadow on the observer sky}
\label{sec:observer-sky}

A static observer at \(r=r_{\rm o}\) carries the orthonormal frame
\begin{equation}
e_{\hat t}
=\frac{1}{\sqrt{A_{\rm o}}}\partial_{t},
\qquad
e_{\hat r}
=\frac{1}{\sqrt{B_{\rm o}}}\partial_{r},
\qquad
e_{\hat\phi}
=\frac{1}{\sqrt{C_{\rm o}}}\partial_{\phi},
\label{eq_tetrad}
\end{equation}
where a subscript \({\rm o}\) denotes evaluation at the observer. The locally
measured temporal and azimuthal photon momenta are
\begin{equation}
p^{\hat t}=\frac{E}{\sqrt{A_{\rm o}}},
\qquad
p^{\hat\phi}=\frac{L}{\sqrt{C_{\rm o}}}.
\label{eq_local-momenta}
\end{equation}
If \(\alpha\) is the angle from the inward radial direction, we choose the
inward-sky branch \(0\leq\alpha\leq\pi/2\), appropriate for a static observer
outside the controlling minimum. The local null cone then gives
\begin{equation}
\sin\alpha
=\frac{|p^{\hat\phi}|}{p^{\hat t}}
=|b|\sqrt{\frac{A_{\rm o}}{C_{\rm o}}}
=\frac{|b|}{\mathcal{R}_{\rm o}}.
\label{eq_local-angle}
\end{equation}
This tetrad relation is the standard arbitrary-distance shadow formula for a
static spherical spacetime \cite{Synge:1966okc,Perlick:2021aok}.
The angular radius of the shadow is obtained by setting
\(|b|=b_{\rm sh}\). This gives
\begin{equation}
\sin^{2}\alpha_{\rm sh}
=\frac{\mathcal{S}_{*}}{\mathcal{S}(r_{\rm o})}
=\frac{\mathcal{A}_{*}}{\mathcal{A}_{\rm opt}(r_{\rm o})}
.
\label{eq_finite-area-law}
\end{equation}
This finite-distance area law is the directly observable form of OAMM.

Spherical symmetry makes the silhouette circular. On a unit celestial disk,
one may write
\begin{equation}
X=\sin\alpha_{\rm sh}\cos\psi,
\qquad
Y=\sin\alpha_{\rm sh}\sin\psi,
\qquad
0\leq\psi<2\pi,
\label{eq_celestial-coordinates}
\end{equation}
and hence
\begin{equation}
X^{2}+Y^{2}
=\frac{\mathcal{S}_{*}}{\mathcal{S}(r_{\rm o})}.
\label{eq_circle}
\end{equation}
The angular diameter and solid angle are
\begin{equation}
\Theta_{\rm sh}=2\alpha_{\rm sh},
\qquad
\Omega_{\rm sh}=2\pi\left(1-\cos\alpha_{\rm sh}\right).
\label{eq_diameter-solid-angle}
\end{equation}

For a distant observer in an asymptotically flat spacetime,
\(A_{\rm o}=1+\mathcal{O}(r_{\rm o}^{-1})\) and
\(C_{\rm o}=r_{\rm o}^{2}+\mathcal{O}(r_{\rm o})\). Therefore
\begin{equation}
\alpha_{\rm sh}
=\frac{b_{\rm sh}}{r_{\rm o}}
+\mathcal{O}(r_{\rm o}^{-2}),
\qquad
D_{\rm sh}=2b_{\rm sh}.
\label{eq_asymptotic-shadow}
\end{equation}
Here \(D_{\rm sh}\) denotes the asymptotic linear diameter in the
impact-parameter plane. It is distinct from the angular diameter
\(\Theta_{\rm sh}\) measured on the observer's sky.
In a nonasymptotically flat spacetime, \(b_{\rm sh}\) can depend on the
normalization chosen for the static time coordinate. The ratio in
Eq. \eqref{eq_finite-area-law} is invariant under a constant rescaling of
that coordinate and is therefore the preferred observable.

Equations \eqref{eq_central-result} and \eqref{eq_finite-area-law} also make
the role of the metric functions transparent. The silhouette of a static
observer depends on \(A(r)\) and \(C(r)\), whereas \(B(r)\) cancels. Metrics
with the same \(A\) and \(C\) on the same static interval consequently have
the same static-observer silhouette even if their radial proper distances and
critical-ray instability rates differ.

\section{Observer-accessible critical curves and invariant accumulation exponents}
\label{sec:accessible-critical-spectrum}

The global minimum of \(\mathcal S=C/A\) selects the capture-shadow edge,
but it does not exhaust the geometric information carried by additional
nondegenerate local minima. Multiple photon spheres, their escape cones, and
their associated critical curves are established features of strong-field
lensing
\cite{Hasse:2001by,Baines:2023bgm,Amo:2023escape,
Guo:2022muy,Guo:2022ghl}. Our purpose here is more specific. We give an
explicit finite-observer scalar criterion that selects every critical family
accessible from the observer side and package the angular position and local
accumulation exponent of each selected family into one ordered data set. This
is the one-dimensional optical-metric specialization of the global
visibility and escape-cone viewpoints of \cite{Baines:2023bgm,
Amo:2023escape}, rather than a new definition of photon-sphere visibility.

\subsection{Strict-record accessibility}
\label{subsec:strict-record-accessibility}

We restrict this subsection to a smooth connected static interval with a
regular absorbing horizon, vacuum metric-null propagation, and a static
observer outside all candidate minima. Let
\begin{equation}
r_{\rm o}>r_{1}>r_{2}>\cdots>r_{N}>r_{\rm h}
\label{eq:ordered-minima}
\end{equation}
be all isolated nondegenerate local minima of \(\mathcal S\) on the
observer-to-horizon interval, ordered as they are encountered by an initially
inward ray. Define
\begin{equation}
\mathcal S_{0}\equiv\mathcal S(r_{\rm o}),
\qquad
\mathcal S_{i}\equiv\mathcal S(r_i),
\qquad
b_i^2\equiv\mathcal S_i .
\label{eq:barrier-data}
\end{equation}
We say that \(r_i\) is observer-accessible as a distinct critical orbit at
its own impact parameter if the ray with \(b=b_i\) reaches a neighborhood of
\(r_i\) before encountering another turning or critical barrier.

The exact path criterion is
\begin{equation}
\mathcal S(r)>\mathcal S_i
\quad\text{for every}\quad r_i<r\leq r_{\rm o}.
\label{eq:path-accessibility}
\end{equation}
When all smooth local minima on the interval have been enumerated, this is
equivalent to the strict running-record rule
\begin{equation}
i\in\mathscr R_{\rm o}
\quad\Longleftrightarrow\quad
\mathcal S_i<
\min\!\left\{\mathcal S_0,\mathcal S_1,\ldots,
\mathcal S_{i-1}\right\}.
\label{eq:record-rule}
\end{equation}
For \(i=1\), the set on the right-hand side contains only
\(\mathcal S_0\).

To prove Eq.~\eqref{eq:path-accessibility}, set \(b^2=\mathcal S_i\) in
the optical radial equation,
\begin{equation}
\left(\frac{d\sigma}{dt}\right)^2
=1-\frac{\mathcal S_i}{\mathcal S(\sigma)},
\qquad
d\sigma=\sqrt{\frac{B}{A}}\,dr.
\label{eq:record-radial-equation}
\end{equation}
The right-hand side must be positive at every intervening section and has a
quadratic zero at the nondegenerate minimum \(r_i\). If an outer section has
\(\mathcal S<\mathcal S_i\), the ray is excluded from that region; if an
outer minimum has \(\mathcal S=\mathcal S_i\), the ray asymptotically
approaches that outer minimum instead. Consequently, equal-depth minima are
not two observer-facing critical generators at the same impact parameter:
the outermost tied minimum screens every inner tied minimum. This strict tie
rule is the reason for the inequality in Eq.~\eqref{eq:record-rule}.

Every member of \(\mathscr R_{\rm o}\) defines a critical circle on the
inward observer sky,
\begin{equation}
\sin^2\alpha_i
=\frac{\mathcal S_i}{\mathcal S_0}.
\label{eq:accessible-critical-angle}
\end{equation}
The record values and their critical angles strictly decrease while moving
inward. Provided that the observer-to-horizon infimum is attained at one of
these minima, the deepest record gives the capture-shadow edge; preceding
records give secondary geometric critical circles outside the shadow. A
non-record minimum may be crossed by rays with smaller \(b\), but it cannot
be reached critically at its own value \(b^2=\mathcal S_i\) from this
observer. If an endpoint infimum is lower than every interior minimum, the
endpoint controls capture and no interior logarithmic critical orbit need
generate the shadow edge.

\subsection{Optical-area Hessian and the invariant exponent}
\label{subsec:optical-hessian-spectrum}

At each accessible minimum, define the dimensionless optical-area Hessian
\begin{equation}
\mathcal H_i
\equiv
\frac{1}{2}
\left.\frac{d^2\mathcal S}{d\sigma^2}\right|_{i}
=\frac{1}{8\pi}
\left.\frac{d^2\mathcal A_{\rm opt}}{d\sigma^2}\right|_{i}>0.
\label{eq:optical-area-hessian}
\end{equation}
Since \(\mathcal R^2=\mathcal S\), stationarity implies
\begin{equation}
\lambda_i^2
=\frac{\mathcal H_i}{\mathcal S_i},
\qquad
\Omega_i
=\left.\frac{d\phi}{dt}\right|_i
=\frac{1}{b_i},
\qquad
K_{{\rm opt},i}=-\lambda_i^2,
\label{eq:hessian-local-relations}
\end{equation}
where \(K_{\rm opt}\) is the Gaussian curvature of the equatorial optical
metric. The natural invariant instability exponent per unit azimuth is
therefore
\begin{equation}
\Xi_i
\equiv b_i\lambda_i
=\frac{\lambda_i}{\Omega_i}
=\sqrt{\mathcal H_i}.
\label{eq:Xi-spectrum-definition}
\end{equation}
Although \(b_i\) and the coordinate-time Lyapunov exponent \(\lambda_i\)
separately change under a constant normalization of the static time
coordinate, their product \(\Xi_i\) does not. It is also invariant under a
smooth monotonic relabeling of the radial coordinate. In the original radial
coordinate,
\begin{equation}
\Xi_i^2
=\left.\frac{A\mathcal S''}{2B}\right|_i
=\left.
\frac{AC''-CA''}{2AB}
\right|_i .
\label{eq:Xi-metric-functions}
\end{equation}
The stationarity condition \(AC'-CA'=0\) was used in the last equality.

The observer-dependent invariant critical-curve spectrum can now be written
as
\begin{equation}
\mathfrak C_{\rm o}
=\left\{
\left(\sin\alpha_i,\Xi_i\right):
i\in\mathscr R_{\rm o}
\right\}.
\label{eq:critical-curve-spectrum}
\end{equation}
The first entry fixes the location of the critical circle on the finite
observer sky, while the second fixes the local rate at which near-critical
trajectories accumulate around it. In particular, the screen locations
depend only on \(A\) and \(C\), whereas \(\Xi_i\) also probes \(B\).

\subsection{Logarithmic winding and angular accumulation}
\label{subsec:ring-accumulation}

Let \(x=\sigma-\sigma_i\) and
\(\varepsilon_i=b^2/b_i^2-1\). Near an isolated nondegenerate minimum,
\begin{equation}
\mathcal S(\sigma)
=b_i^2+\Xi_i^2x^2+\mathcal O(x^3).
\label{eq:local-S-Xi}
\end{equation}
A complete local reflection or transmission through the near-critical
neighborhood then gives the standard logarithmic contribution
\begin{equation}
\Delta\phi_i(b)
=-\frac{1}{\Xi_i}
\ln\left|\frac{b^2}{b_i^2}-1\right|
+\mathcal B_i^{\pm}+o(1),
\label{eq:logarithmic-winding}
\end{equation}
where \(\mathcal B_i^{\pm}\) contains the regular, source- and
observer-dependent part of the trajectory. Thus the usual strong-deflection
coefficient is
\begin{equation}
\bar a_i=\frac{1}{\Xi_i}
=\sqrt{\frac{2A_iB_i}{A_iC_i''-C_iA_i''}} .
\label{eq:strong-deflection-Xi}
\end{equation}
This relation is consistent with the established connection among the
strong-deflection coefficient, orbital frequency, and Lyapunov exponent
\cite{Bozza:2002zj,Stefanov:2010xz}.

For a specified source--observer family whose successive images differ by
an asymptotic azimuthal step \(\Delta\phi_{\rm step}\), Eq.~\eqref{eq:logarithmic-winding}
implies
\begin{equation}
\lim_{n\to\infty}
\frac{|b_{i,n+1}-b_i|}{|b_{i,n}-b_i|}
=\exp\!\left(-\Xi_i\Delta\phi_{\rm step}\right).
\label{eq:general-accumulation-ratio}
\end{equation}
For same-parity images separated by one full winding,
\(\Delta\phi_{\rm step}=2\pi\), whereas half-orbit indexing gives
\(\Delta\phi_{\rm step}=\pi\). At a finite static observer,
\(b=\sqrt{\mathcal S_0}\sin\alpha\), so the same limiting ratio holds for
the angular separations \(|\alpha_{i,n}-\alpha_i|\). This result concerns
asymptotic positions, gaps, and widths; it does not imply a universal flux
ratio. Actual visibility and brightness also require a source distribution,
transparent propagation, and a lens equation that admits the relevant
repeated-winding trajectories
\cite{Johnson:2019photonring}.

For orientation, Table~\ref{tab:representative-Xi-spectrum} lists the
invariant exponents for the representative parameters used in
Sec.~\ref{sec:applications}. The last column adopts the same-parity,
full-winding convention \(\Delta\phi_{\rm step}=2\pi\). These values are
not an equal-parameter comparison because the deformation parameters have
different physical meanings.

\begin{table}[t]
\caption{Invariant instability and strong-winding data for the
representative models used in Sec.~\ref{sec:applications}.}
\label{tab:representative-Xi-spectrum}
\centering
\begin{tabular}{lccc}
\hline\hline
Geometry and parameter
& \(\Xi_*\)
& \(\bar a_*=1/\Xi_*\)
& \(\exp(-2\pi\Xi_*)\) \\
\hline
Reissner--Nordstr\"om, \(Q/M=0.8\)
& \(0.890338\) & \(1.123169\) & \(3.7195\times10^{-3}\) \\
Bardeen, \(g/M=0.5\)
& \(0.917653\) & \(1.089736\) & \(3.1329\times10^{-3}\) \\
Charged dilaton, \(a/M=1\)
& \(0.861022\) & \(1.161411\) & \(4.4718\times10^{-3}\) \\
Kottler, \(\Lambda M^2=10^{-3}\)
& \(1.000000\) & \(1.000000\) & \(1.8674\times10^{-3}\) \\
\hline\hline
\end{tabular}
\end{table}

The table makes explicit that the critical-circle scale and the local
accumulation rate are complementary observables. In particular, models with
similar shadow sizes need not have similar \(\Xi_*\), while changing only
the radial metric function \(B\) leaves the static critical-circle angle
unchanged but changes both \(\Xi_*\) and the asymptotic ring spacing.

\subsection{A record-set transition between two optical barriers}
\label{subsec:record-set-transition}

The strict-record formulation predicts an observer-dependent transition that
is not visible from a list of stationary radii alone. Consider two minima,
an outer one \(r_1\) and an inner one \(r_2\), with
\(\mathcal S_1<\mathcal S_0\), and let a deformation parameter change their
relative depths. If no lower endpoint or additional minimum intervenes, then
\begin{equation}
\mathscr R_{\rm o}
=
\begin{cases}
\{1,2\}, & \mathcal S_2<\mathcal S_1,\\[1mm]
\{1\}, & \mathcal S_2\geq\mathcal S_1.
\end{cases}
\label{eq:two-record-transition}
\end{equation}
For \(\mathcal S_2<\mathcal S_1\), the outer minimum produces a secondary
critical circle and the inner minimum produces the smaller capture-shadow
edge. At equality, the outer separatrix screens the tied inner orbit. After
the ordering reverses, the outer minimum alone is critically accessible and
also controls capture. The lower envelope
\begin{equation}
b_{\rm sh}^2
=\min\{\mathcal S_1,\mathcal S_2\}
\label{eq:record-lower-envelope}
\end{equation}
is continuous at a transverse crossing but is generically nondifferentiable,
while the number of observer-accessible critical families changes from two
to one. Near an almost degenerate pair, the regular term
\(\mathcal B_i^{\pm}\) can become large because the inner critical ray passes
quasi-critically through the outer barrier. Therefore two independent
single-barrier asymptotic expansions should not be assumed to remain uniform
through the crossing.

The logarithmic construction above requires isolated smooth minima with
\(\mathcal S_{,\sigma\sigma}>0\). Degenerate minima instead lead to a
higher-order, generally power-law critical behavior; stable optical maxima
describe trapping rather than the unstable logarithmic family. Endpoint
infima, cusps, thin shells, observers between barriers, and initially outward
rays require their corresponding one-sided or two-sided accessibility
analysis and are outside the strict inward-record theorem stated here.

\section{Perturbative deformations}
\label{sec:perturbations}

The minimum formulation is especially efficient for a metric that is a small
deformation of a known geometry. The expansion below is the vacuum-null,
single-function specialization of more general perturbative shadow
frameworks \cite{Kobialko:2024zhc,Pantig:2025deu}. Let
\begin{equation}
\mathcal{S}(r,\epsilon)
=\mathcal{S}_{0}(r)
+\epsilon\mathcal{S}_{1}(r)
+\epsilon^{2}\mathcal{S}_{2}(r)
+\mathcal{O}(\epsilon^{3}),
\label{eq_S-perturbation}
\end{equation}
and suppose the undeformed minimum \(r_{0}\) satisfies
\begin{equation}
\mathcal{S}_{0}'(r_{0})=0,
\qquad
\mathcal{S}_{0}''(r_{0})>0.
\label{eq_undeformed-minimum}
\end{equation}
These local conditions identify a smooth stationary branch but do not by
themselves make it the perturbation of the global capture threshold. For the
single-branch expansion below, we additionally assume that \(r_{0}\) is an
isolated strict global minimum on the undeformed observer-accessible interval
and that its value is separated by a positive gap from every other
stationary candidate and relevant endpoint limit. We also restrict
\(\epsilon\) to a range in which this continued branch remains admissible and
controlling. These assumptions exclude a branch switch within the expansion
range.
Write the shifted critical radius as
\begin{equation}
r_{*}
=r_{0}+\epsilon r_{1}+\epsilon^{2}r_{2}
+\mathcal{O}(\epsilon^{3}).
\label{eq_r-perturbation}
\end{equation}
Expanding \(\mathcal{S}'(r_{*},\epsilon)=0\) to first order gives
\begin{equation}
r_{1}
=-\frac{\mathcal{S}_{1}'(r_{0})}
{\mathcal{S}_{0}''(r_{0})}
.
\label{eq_r1}
\end{equation}

More importantly, the value of the minimum through second order is
\begin{equation}
\mathcal{S}_{*}
=\mathcal{S}_{0}
+\epsilon\mathcal{S}_{1}
+\epsilon^{2}
\left[
\mathcal{S}_{2}
-\frac{(\mathcal{S}_{1}')^{2}}
{2\mathcal{S}_{0}''}
\right]
+\mathcal{O}(\epsilon^{3})
,
\label{eq_S-min-perturbation}
\end{equation}
where every quantity on the right-hand side is evaluated at \(r=r_{0}\).
The term \(r_{1}\mathcal{S}_{0}'(r_{0})\) vanishes, so the first-order shadow
correction does not require the first-order displacement of the critical
radius.

Defining \(b_{0}=\sqrt{\mathcal{S}_{0}(r_{0})}\), one obtains
\begin{equation}
\begin{aligned}
b_{\rm sh}
={}&b_{0}
+\epsilon\frac{\mathcal{S}_{1}}{2b_{0}}
\\
&+\epsilon^{2}
\left[
\frac{\mathcal{S}_{2}}{2b_{0}}
-\frac{(\mathcal{S}_{1}')^{2}}
{4b_{0}\mathcal{S}_{0}''}
-\frac{\mathcal{S}_{1}^{2}}{8b_{0}^{3}}
\right]
+\mathcal{O}(\epsilon^{3}).
\end{aligned}
\label{eq_b-perturbation}
\end{equation}
Thus the first-order change in the shadow radius is obtained simply by
evaluating the first-order change in \(C/A\) at the undeformed critical
radius. The displacement \(r_{1}\) becomes necessary only when the location
of the critical orbit is itself required or when second-order corrections are
retained.

If the positive-gap assumption fails, no single branch expansion need equal
the global threshold. Instead, each admissible stationary branch and each
relevant endpoint must be expanded separately, including the displacement of
a horizon or observer endpoint when present, and the result selected
piecewise:
\begin{equation}
\mathcal{S}_{*}(\epsilon)
=\min\!\left(
\left\{\mathcal{S}_{*,i}(\epsilon)\right\}_{i},
\left\{\lim_{r\to r_{e}(\epsilon)}
\mathcal{S}(r,\epsilon)\right\}_{e}
\right).
\label{eq_piecewise-perturbative-minimum}
\end{equation}
At a crossing of two controlling values,
\(\mathcal{S}_{*}(\epsilon)\) is generally only piecewise differentiable, so
Eqs. \eqref{eq_S-min-perturbation} and \eqref{eq_b-perturbation} apply to the
individual branches rather than across the switch.

\section{Prescription and domain of validity}
\label{sec:scope}

For a static spherical black hole, the calculation can be summarized by the
following steps.
\begin{enumerate}
\item Identify \(A(r)\), \(B(r)\), and \(C(r)\), together with the connected
static interval from the observer to the black hole horizon.
\item Form \(\mathcal{S}(r)=C(r)/A(r)\).
\item Find every admissible interior solution of
\(AC'-CA'=0\), include any relevant endpoint limits, and compare their
values of \(\mathcal{S}\).
\item Set \(b_{\rm sh}=\sqrt{\mathcal{S}_{*}}\), where
\(\mathcal{S}_{*}\) is the global infimum on the accessible interval.
\item For a static observer, compute
\(\alpha_{\rm sh}=\arcsin\sqrt{\mathcal{S}_{*}/\mathcal{S}_{\rm o}}\) on
the inward-sky branch \(0\leq\alpha_{\rm sh}\leq\pi/2\).
\end{enumerate}
No numerical integration of photon trajectories is required to locate the
shadow edge.

The present construction assumes a static, spherically symmetric spacetime,
vacuum photon propagation, and a static observer outside the controlling
minimum. Rotation replaces the single circular silhouette by a
direction-dependent curve, so a single spherical area cannot encode the full
shadow. A dispersive plasma introduces frequency dependence and requires an
appropriate frequency-dependent optical geometry
\cite{Perlick:2015vta}. A moving observer sees an
aberrated image even though the underlying capture set is unchanged.

The divergence in Eq. \eqref{eq_asymptotic-limits} is sufficient but not
necessary. If the smallest value of \(C/A\) is approached at a boundary, the
infimum in Eq. \eqref{eq_central-result} remains meaningful, although the
interior stationarity condition \eqref{eq_stationarity} and the circular-orbit
interpretation of the equality case do not apply. Likewise, a degenerate
interior minimum determines a threshold but lies outside the quadratic
Lyapunov analysis of Eqs. \eqref{eq_local-instability}--
\eqref{eq_lambda-coordinate}.
Observers located at or inside the controlling photon sphere also require
care with the choice of radial direction and the branch of the local sky
angle. The simple range \(0\leq\alpha_{\rm sh}\leq\pi/2\) used here assumes
\(r_{\rm o}>r_{*}\).

Finally, the construction determines the geometric capture silhouette, not
an intensity map. Emission, absorption, plasma dynamics, and radiative
transfer determine the brightness distribution around that boundary.

\section{Implementation of the optical-area minimum method}
\label{sec:applications}

\subsection{Synthetic profile with two exterior minima}
\label{subsec:synthetic}

To display the global selection step directly, consider the deliberately
synthetic asymptotically flat metric
\begin{equation}
A(x)=\frac{x^{2}}{s(x)},
\qquad
B(x)=\frac{1}{A(x)},
\qquad
C(x)=M^{2}x^{2},
\qquad
x=\frac{r}{M}>1,
\label{eq_synthetic-functions}
\end{equation}
where
\begin{equation}
\begin{aligned}
s(x)\equiv\frac{\mathcal{S}(x)}{M^{2}}
={}&384\ln 2-89+x^{2}+2x-384\ln x
\\
&+294\ln(x-1)+\frac{108}{x-1}.
\end{aligned}
\label{eq_synthetic-S}
\end{equation}
This profile is an algorithmic illustration rather than a proposed matter
model. It has a horizon at \(x=1\), where
\(s(x)\sim108/(x-1)\) and \(A(x)\to0\), while
\(s(x)=x^{2}+2x+\mathcal{O}(\ln x)\) gives \(A(x)\to1\) at infinity.
Its derivative factorizes as
\begin{equation}
s'(x)
=\frac{2(x+8)(x-2)(x-3)(x-4)}{x(x-1)^{2}}.
\label{eq_synthetic-S-prime}
\end{equation}
Thus an observer at \(x_{\rm o}=6\) has two admissible exterior minima,
separated by a maximum. All stationary candidates and accessible endpoint
values are listed in Table \ref{tab:synthetic-candidates}.

\begin{table}[t]
\caption{Global candidate comparison for the synthetic profile
\eqref{eq_synthetic-S} on \(1<x\leq6\).}
\label{tab:synthetic-candidates}
\begin{ruledtabular}
\begin{tabular}{lcc}
Location & Classification & \(\mathcal{S}/M^{2}\) \\
\hline
\(x\to1^{+}\) & horizon endpoint & \(+\infty\) \\
\(x=2\) & local minimum & \(27.000000\) \\
\(x=3\) & local maximum & \(28.086670\) \\
\(x=4\) & local minimum & \(27.823496\) \\
\(x=6\) & observer endpoint & \(31.907627\)
\end{tabular}
\end{ruledtabular}
\end{table}

The inner minimum at \(x=2\), rather than the outer minimum at \(x=4\), is
therefore controlling:
\begin{equation}
r_{*}=2M,
\qquad
b_{\rm sh}=3\sqrt{3}\,M.
\label{eq_synthetic-result}
\end{equation}
This example demonstrates why solving the local stationarity equation alone
is insufficient: every admissible candidate and endpoint value must be
compared before the capture threshold is assigned.

\subsection{Reissner--Nordstr\"om spacetime}
\label{subsec:rn}

For the Reissner--Nordstr\"om geometry
\cite{Perlick:2021aok,EventHorizonTelescope:2021dqv},
\begin{equation}
A(r)
=1-\frac{2M}{r}+\frac{Q^{2}}{r^{2}},
\qquad
B(r)=\frac{1}{A(r)},
\qquad
C(r)=r^{2}.
\label{eq_rn-metric-functions}
\end{equation}
For \(|Q|\leq M\), the outer horizon is
\begin{equation}
r_{+}=M+\sqrt{M^{2}-Q^{2}}.
\label{eq_rn-horizon}
\end{equation}
The optical-area function and its derivative are
\begin{equation}
\mathcal{S}(r)
=\frac{r^{4}}{r^{2}-2Mr+Q^{2}},
\label{eq_rn-S}
\end{equation}
\begin{equation}
\mathcal{S}'(r)
=\frac{2r^{3}(r^{2}-3Mr+2Q^{2})}
{(r^{2}-2Mr+Q^{2})^{2}}.
\label{eq_rn-S-prime}
\end{equation}
The stationary radii are
\begin{equation}
r_{\pm}^{\rm stat}
=\frac{3M\pm\sqrt{9M^{2}-8Q^{2}}}{2}.
\label{eq_rn-stationary}
\end{equation}
Comparison with the exterior static interval selects the larger root,
\begin{equation}
r_{*}
=\frac{3M+\sqrt{9M^{2}-8Q^{2}}}{2}
.
\label{eq_rn-rstar}
\end{equation}
Using \(r_{*}^{2}-3Mr_{*}+2Q^{2}=0\), the shadow radius becomes
\begin{equation}
b_{\rm sh}^{2}
=\frac{r_{*}^{4}}{Mr_{*}-Q^{2}}
.
\label{eq_rn-b}
\end{equation}
The finite-distance angular radius is
\begin{equation}
\sin\alpha_{\rm sh}
=\frac{b_{\rm sh}}{r_{\rm o}}
\sqrt{1-\frac{2M}{r_{\rm o}}+\frac{Q^{2}}{r_{\rm o}^{2}}}.
\label{eq_rn-angle}
\end{equation}
For \(Q=0\), Eqs. \eqref{eq_rn-rstar} and \eqref{eq_rn-b} reduce to
\(r_{*}=3M\) and \(b_{\rm sh}=3\sqrt{3}M\). For the representative choice
\(Q=0.8M\),
\begin{equation}
\frac{r_{*}}{M}=2.484885780,
\qquad
\frac{b_{\rm sh}}{M}=4.545986272.
\label{eq_rn-numerics}
\end{equation}
A static observer at \(r_{\rm o}=10M\) measures
\begin{equation}
\alpha_{\rm sh}=24.0936^{\circ},
\qquad
\Theta_{\rm sh}=48.1872^{\circ}.
\label{eq_rn-angle-numerics}
\end{equation}

\subsection{Bardeen black hole}
\label{subsec:bardeen}

The Bardeen regular black hole, interpreted as a nonlinear magnetic monopole
solution \cite{Ayon-Beato:1998hmi,Ayon-Beato:2000mjt}, is described by
\begin{equation}
A(r)
=1-\frac{2Mr^{2}}{(r^{2}+g^{2})^{3/2}},
\qquad
B(r)=\frac{1}{A(r)},
\qquad
C(r)=r^{2},
\label{eq_bardeen-functions}
\end{equation}
where \(g\) is the magnetic-charge parameter. A horizon exists in the
black hole domain
\begin{equation}
\frac{|g|}{M}\leq\frac{4}{3\sqrt{3}}.
\label{eq_bardeen-bound}
\end{equation}
The outer horizon, defined as the largest positive root of \(A(r)=0\), is the
lower endpoint of the exterior interval used in the minimization. Outside
the bound in Eq. \eqref{eq_bardeen-bound}, the same metric is horizonless and
the observer-to-horizon shadow prescription does not apply as written. Since
\begin{equation}
A'(r)
=\frac{2Mr(r^{2}-2g^{2})}{(r^{2}+g^{2})^{5/2}},
\label{eq_bardeen-A-prime}
\end{equation}
the interior-minimum condition \(rA'-2A=0\) reduces to
\begin{equation}
3Mr_{*}^{4}
=(r_{*}^{2}+g^{2})^{5/2}
.
\label{eq_bardeen-rstar}
\end{equation}
If several positive solutions exist, only those in the exterior static
region are admissible, and their values of \(\mathcal{S}\) must be compared.
For the black hole parameters used below, the outer admissible solution is
the global exterior minimum.

Equation \eqref{eq_bardeen-rstar} also simplifies the metric function at the
critical radius. One obtains
\begin{equation}
A_{*}
=\frac{r_{*}^{2}-2g^{2}}{3r_{*}^{2}}.
\label{eq_bardeen-Astar}
\end{equation}
Consequently,
\begin{equation}
b_{\rm sh}^{2}
=\frac{3r_{*}^{4}}{r_{*}^{2}-2g^{2}}
.
\label{eq_bardeen-b}
\end{equation}
For \(g=0.5M\), which lies within the black hole domain, numerical solution of
Eq. \eqref{eq_bardeen-rstar} gives
\begin{equation}
\frac{r_{*}}{M}=2.768711993,
\qquad
\frac{b_{\rm sh}}{M}=4.960036554.
\label{eq_bardeen-numerics}
\end{equation}
At \(r_{\rm o}=10M\), the finite-distance result is
\begin{equation}
\alpha_{\rm sh}=26.3496^{\circ},
\qquad
\Theta_{\rm sh}=52.6991^{\circ}.
\label{eq_bardeen-angle}
\end{equation}
This shadow radius is approximately \(4.54\%\) smaller than the Schwarzschild
value for the same mass parameter. Shadows of rotating regular black holes and
EHT-motivated constraints on magnetically charged nonlinear-electrodynamic
geometries provide complementary studies of this model class
\cite{Abdujabbarov:2016hnw,Amir:2016cen,Allahyari:2019jqz}.

\subsection{Charged dilaton black hole with a nonareal radial coordinate}
\label{subsec:dilaton}

Consider the electrically charged dilaton black hole in the Einstein frame
\cite{Garfinkle:1990qj}, with line element specified by
\begin{equation}
A(r)=1-\frac{2M}{r},
\qquad
B(r)=\frac{1}{A(r)},
\qquad
C(r)=r(r-a),
\qquad
0\leq a<2M.
\label{eq_dilaton-functions}
\end{equation}
In the standard normalization of the Garfinkle--Horowitz--Strominger
solution, \(a=Q_{\rm e}^{2}/M\), where \(Q_{\rm e}\) is the electric charge.
The bound above is the black hole condition in this parametrization.
The horizon is at \(r_{\rm h}=2M\), while the surface \(r=a\) lies inside it.
The coordinate \(r\) is not an areal radius because the physical area of a
spacetime sphere is \(4\pi r(r-a)\).

The optical-area function is
\begin{equation}
\mathcal{S}(r)
=\frac{r^{2}(r-a)}{r-2M},
\label{eq_dilaton-S}
\end{equation}
with derivative
\begin{equation}
\mathcal{S}'(r)
=\frac{r[2r^{2}-(6M+a)r+4aM]}{(r-2M)^{2}}.
\label{eq_dilaton-S-prime}
\end{equation}
The exterior minimum occurs at
\begin{equation}
r_{*}
=\frac{6M+a+\sqrt{36M^{2}-20aM+a^{2}}}{4}
,
\label{eq_dilaton-rstar}
\end{equation}
and the shadow radius is
\begin{equation}
b_{\rm sh}^{2}
=\frac{r_{*}^{2}(r_{*}-a)}{r_{*}-2M}
.
\label{eq_dilaton-b}
\end{equation}
For \(a=M\),
\begin{equation}
\frac{r_{*}}{M}
=\frac{7+\sqrt{17}}{4}
=2.780776406,
\qquad
\frac{b_{\rm sh}}{M}=4.199595154.
\label{eq_dilaton-numerics}
\end{equation}
At \(r_{\rm o}=10M\), one has
\(\mathcal{R}_{\rm o}/M=\sqrt{90/0.8}=10.60660172\), and therefore
\begin{equation}
\alpha_{\rm sh}=23.3247^{\circ},
\qquad
\Theta_{\rm sh}=46.6494^{\circ}.
\label{eq_dilaton-angle}
\end{equation}
Replacing \(C(r)\) by \(r^{2}\) would give an incorrect result. This example
shows why the full angular metric function, rather than the radial coordinate
itself, must be used in the optical area.

The radial-coordinate invariance can be checked explicitly by introducing
the areal coordinate
\begin{equation}
R=\sqrt{r(r-a)},
\qquad
r(R)=\frac{a+\sqrt{a^{2}+4R^{2}}}{2}.
\label{eq_dilaton-areal-coordinate}
\end{equation}
In this coordinate the metric functions are
\begin{equation}
\widetilde A(R)=1-\frac{2M}{r(R)},
\qquad
\widetilde B(R)
=\frac{4R^{2}}
{\left(a^{2}+4R^{2}\right)\widetilde A(R)},
\qquad
\widetilde C(R)=R^{2}.
\label{eq_dilaton-areal-functions}
\end{equation}
Consequently,
\begin{equation}
\widetilde{\mathcal S}(R)
=\frac{R^{2}}{\widetilde A(R)}
=\mathcal S\bigl(r(R)\bigr).
\label{eq_dilaton-coordinate-invariance}
\end{equation}
Thus the minimum is relabeled as
\(R_{*}=\sqrt{r_{*}(r_{*}-a)}\), while its value
\(\mathcal S_{*}\) is unchanged. The observer is likewise relabeled by
\(R_{\rm o}=\sqrt{r_{\rm o}(r_{\rm o}-a)}\), so the ratio in
Eq. \eqref{eq_finite-area-law} and the angle in
Eq. \eqref{eq_dilaton-angle} are identical in the two radial coordinates.
Strong-deflection lensing and shadow observables have also been studied for
rotating dilaton--axion extensions of this solution
\cite{Gyulchev:2006zg,Wei:2013kza}.

\subsection{Kottler black hole}
\label{subsec:kottler}

The Kottler, or Schwarzschild--de Sitter, geometry
\cite{Stuchlik:1999qk,Gibbons:2008ru} has
\begin{equation}
A(r)
=1-\frac{2M}{r}-\frac{\Lambda r^{2}}{3},
\qquad
B(r)=\frac{1}{A(r)},
\qquad
C(r)=r^{2}.
\label{eq_kottler-functions}
\end{equation}
For \(0<9\Lambda M^{2}<1\), the black hole and cosmological horizons bound a
connected static region. We place the observer in that region with
\(r_{\rm o}>3M\). The optical-area function satisfies
\begin{equation}
\mathcal{S}(r)=\frac{r^{2}}{A(r)},
\qquad
\mathcal{S}'(r)=\frac{2(r-3M)}{A^{2}(r)}.
\label{eq_kottler-S}
\end{equation}
Thus the cosmological constant cancels from the stationarity equation, and
\begin{equation}
r_{*}=3M.
\label{eq_kottler-rstar}
\end{equation}
It nevertheless changes the minimum value. The result is
\begin{equation}
b_{\rm sh}^{2}
=\mathcal{S}_{*}
=\frac{27M^{2}}{1-9\Lambda M^{2}}
.
\label{eq_kottler-b}
\end{equation}
For a finite static observer, the invariant angular result is
\begin{equation}
\sin^{2}\alpha_{\rm sh}
=\frac{27M^{2}}{1-9\Lambda M^{2}}
\frac{1-\dfrac{2M}{r_{\rm o}}-\dfrac{\Lambda r_{\rm o}^{2}}{3}}
{r_{\rm o}^{2}}
.
\label{eq_kottler-angle}
\end{equation}
This example separates the critical radius from the observed shadow size.
\(\Lambda\) leaves \(r_{*}\) unchanged but affects both the minimum optical
area and the local conversion from impact parameter to angle.

For \(\Lambda M^{2}=10^{-3}\) and \(r_{\rm o}=10M\),
\begin{equation}
\frac{b_{\rm sh}}{M}=5.219694135,
\qquad
\alpha_{\rm sh}=27.1959^{\circ},
\qquad
\Theta_{\rm sh}=54.3917^{\circ}.
\label{eq_kottler-numerics}
\end{equation}
Because there is no asymptotically flat observer in the de Sitter static
region, Eq. \eqref{eq_kottler-angle}, rather than a Euclidean screen at
infinity, is the physically relevant comparison
\cite{Perlick:2018iye,Perlick:2021aok}. The value of \(b_{\rm sh}\) quoted
in Eq. \eqref{eq_kottler-numerics} uses the time normalization displayed in
Eq. \eqref{eq_kottler-functions}. The locally measured angle is unchanged by
a constant rescaling of that coordinate.

For reference, the representative finite-distance results are collected in
Table \ref{tab:applications}. The deformation parameters have different
physical meanings. The table illustrates the common calculation rather than
an equal-parameter comparison.

\begin{table}[t]
\caption{Representative results for a static observer at
\(r_{\rm o}=10M\). The table illustrates the common calculation rather than
an equal-parameter comparison. For Kottler, \(b_{\rm sh}\) refers to the time
normalization in Eq. \eqref{eq_kottler-functions}. The invariant
finite-distance comparison is the locally measured angular diameter
\(\Theta_{\rm sh}\).}
\label{tab:applications}
\begin{ruledtabular}
\begin{tabular}{lccc}
Spacetime and parameter
& \(r_{*}/M\)
& \(b_{\rm sh}/M\)
& \(\Theta_{\rm sh}\) \\
\hline
Reissner--Nordstr\"om, \(Q/M=0.8\)
& \(2.484886\) & \(4.545986\) & \(48.1872^{\circ}\) \\
Bardeen, \(g/M=0.5\)
& \(2.768712\) & \(4.960037\) & \(52.6991^{\circ}\) \\
Charged dilaton, \(a/M=1\)
& \(2.780776\) & \(4.199595\) & \(46.6494^{\circ}\) \\
Kottler, \(\Lambda M^{2}=10^{-3}\)
& \(3.000000\) & \(5.219694\) & \(54.3917^{\circ}\)
\end{tabular}
\end{ruledtabular}
\end{table}

\section{Bound and rosette null orbits in the optical-area landscape}
\label{sec:bound-rosette}

The analysis above has focused mainly on local minima of $\mathcal S=\frac{C}{A}$, because such minima generate unstable circular null orbits and determine capture and critical ray accumulation. Local maxima of $\mathcal S$, however, have a complementary dynamical meaning. They correspond to stable circular null orbits, often called stable photon spheres or anti-photon spheres \cite{Cvetic:2016bxi}. Stable photon trapping and multiple-photon sphere configurations are known to occur in static spherical geometries, including black hole solutions, and can be associated with long-lived wave modes \cite{Guo:2022muy,Guo:2022ghl}.

Related distinctions between circular, bounded, and closed geodesic motion have also appeared in studies of Kerr geodesics~\cite{Grib:2014negative,Vertogradov:2015negative}. We distinguish between bound noncircular null orbits, closed resonant null orbits, and nonclosed rosette orbits. Periodic relativistic orbits can naturally be classified by rational ratios of their fundamental frequencies \cite{Levin:2008Periodic}.

\subsection{Bound motion around an optical-area maximum}

Consider a local maximum $r=r_{s}$ of the optical-area function,
\begin{equation} \label{eq:stable1}
\mathcal S'(r_{s})=0,\qquad\mathcal S''(r_{s})<0,
\end{equation}
and suppose that it lies between two neighboring local minima $r_{L}<r_{s}<r_{R}$. 
Since $r_{s}$ is a local maximum, $\mathcal S(r_{s})>\mathcal S(r_{L})$ and $\mathcal S(r_{s})>\mathcal S(r_{R})$.

The radial equation \eqref{eq_radial-crossing} shows that null motion is allowed where
\begin{equation}
\mathcal S(r)\geq b^2.
\end{equation}
Therefore, if
\begin{equation} \label{eq:condition1}
\max(\mathcal S_{L},\mathcal S_{R})<b^2<\mathcal S_{s},
\end{equation}
there are two turning points
\begin{equation} \label{eq:condition2}
r_{p}<r_{s}<r_{a}, \qquad\mathcal S(r_{p})=\mathcal S(r_{\rm a})=b^2,
\end{equation}
with
\begin{equation}
r_{L}<r_{p}<r_{s}<r_{a}<r_{R}.
\end{equation}
The interval $r_{p}\leq r\leq r_{a}$ is allowed, whereas the neighboring regions around both optical minima are forbidden at the same impact parameter. The photon is therefore trapped and oscillates between the finite radii $r_{p}$ and $r_{a}$. We refer to such a trajectory as a bound noncircular null orbit.

At the limiting value $b^2=\mathcal S_{s}$, the two turning points merge at $r_{s}$, and the bound family reduces to the stable circular null orbit.

\subsection{Closed resonances and rosette orbits}

For $b>0$, the azimuthal advance during one complete radial oscillation $r_{p}\rightarrow r_{a}\rightarrow r_{p}$ follows directly from the null geodesic equations:
\begin{equation} \label{eq:angle1}
\Delta\phi_r(b) =2\int_{r_{p}}^{r_{a}}\frac{b\sqrt{AB}}{C\sqrt{1-\frac{b^2}{\mathcal S}}}dr.
\end{equation}
It is convenient to define the dimensionless rotation number
\begin{equation}
\nu(b)\equiv\frac{\Delta\phi_r(b)}{2\pi}.
\end{equation}

If
\begin{equation}
\nu(b)=\frac{m}{n},\qquad m,n\in\mathbb N
\end{equation}
the photon returns to its initial spatial position after $n$ radial oscillations and $m$ complete azimuthal revolutions. We call this a closed resonant null orbit. For a generic value of the impact parameter, $\nu(b)$ is irrational, so the apsidal direction never repeats exactly and the trajectory forms a precessing rosette. Thus closed resonances form a special subset of the continuous family of bound null trajectories.

The local optical-area Hessian introduced in Eq.~\eqref{eq:optical-area-hessian} also has a direct interpretation for the stable branch. At $r=r_{s}$
\begin{eqnarray}
\mathcal H_{s}&=&\frac{1}{2}\frac{d^2\mathcal S}{d\sigma^2}|_{s}<0,\nonumber \\
\Upsilon_{s}&\equiv &\sqrt{-\mathcal H_{s}}.
\end{eqnarray}
Let
\begin{equation}
b^2=\mathcal S_{s}-\delta,\qquad 0<\delta\ll\mathcal S_{s},
\end{equation}
and write
\begin{equation}
y=\sigma-\sigma_{s}.
\end{equation}
Near the stable circular orbit,
\begin{equation}
\mathcal S(\sigma)=\mathcal S_{s}+\mathcal H_{s}y^2+\mathcal O(y^3).
\end{equation}
Using Eq.~\eqref{eq:record-radial-equation}, the leading radial equation is
\begin{equation}
\left(\frac{dy}{dt}\right)^2+\frac{-\mathcal H_{s}}{\mathcal S_{s}}y^2=\frac{\delta}{\mathcal S_{s}}.
\end{equation}
Consequently,
\begin{equation}
\frac{d^2y}{dt^2}+\omega_r^2 y=0,\qquad\omega_r^2=-\frac{\mathcal H_{s}}{\mathcal S_{s}}.
\end{equation}
The angular frequency of the limiting circular orbit is
\begin{equation}
\Omega_{s}=\frac{1}{\sqrt{\mathcal S_{s}}},
\end{equation}
and therefore
\begin{equation}
\frac{\omega_r}{\Omega_{s}}=\Upsilon_{s}=\sqrt{-\mathcal H_{s}}.
\end{equation}
Hence the small-amplitude rotation number satisfies
\begin{equation}
\lim_{b^2\to\mathcal S_{s}^{-}}\nu(b)=\frac{1}{\Upsilon_{s}}.
\end{equation}

Equations \eqref{eq:Xi-spectrum-definition} and \eqref{eq:stable-frequency-ratio} show the complementary role of the sign of the optical-area Hessian. A minimum, $\mathcal H>0$, gives the invariant exponential instability parameter $\Xi=\sqrt{\mathcal H}$, whereas a maximum, $\mathcal H<0$, gives the invariant ratio $\Upsilon=\sqrt{-\mathcal H}$ that controls small bound oscillations.

\subsection{Bound and rosette motion in the synthetic two-barrier profile}

The synthetic profile of Sec.~\ref{subsec:synthetic} already contains the required optical structure. From Eqs.~\eqref{eq_synthetic-S} and \eqref{eq_synthetic-S-prime}, the two minima are located at $x=2$ and $x=4$, while $x=3$ is a local maximum. Their values are $s(2)=27,s(3)\approx 28.09, s(4)\approx 27.82$. Introducing $\beta=\frac{b}{M}$, the bound-orbit window is therefore
\begin{equation}
27.82<\beta^2<28.09
\end{equation}

The optical radial coordinate satisfies
\begin{equation}
d\sigma=\frac{dr}{A}.
\end{equation}
At the stationary point $x=3$,
\begin{equation}
\mathcal H_{s}=\frac{1}{2}A^2(3)s''(3)\approx -0.09
\end{equation}
Thus $\Upsilon_{s}\approx 0.31$ and the small-amplitude bound trajectories satisfy
\begin{equation}
\nu_0\equiv\lim_{\beta^2\to s(3)^-}\nu(\beta)=\frac{1}{\Upsilon_{s}}\approx 3.26
\end{equation}

For this metric, Eq.~\eqref{eq:angle1} reduces to
\begin{equation} \label{eq:angle2}
\Delta\phi_r(\beta)=2\int_{x_{p}}^{x_{a}}\frac{\beta}{x^2\sqrt{1-\frac{\beta^2}{s(x)}}}dx,
\end{equation}
where
\begin{equation}
s(x_{p})=s(x_{a})=\beta^2, \qquad 2<x_{p}<3<x_{a}<4.
\end{equation}

As a representative nonresonant example, choosing $\beta^2=28$ gives $x_{p}\approx 2.72, x_{a}\approx 3.36$ and numerical integration of Eq.~\eqref{eq:angle2} yields $\nu=\frac{\Delta\phi_r}{2\pi}\approx 3.39$ This value is not tuned to a low-order rational resonance and represents the generic precessing rosette behavior. Closed resonances can be obtained by tuning the impact parameter. 

In physical multi-photon-sphere black holes such trapping can affect wave
propagation and can generate long-lived modes
\cite{Cvetic:2016bxi,Guo:2022ghl}. A detailed study of the observational
signatures of the resonant and rosette families is beyond the scope of the
present work.

\section{Conclusions}
\label{sec:conclusions}

We have reformulated the shadow calculation for static and spherically
symmetric black holes as a global minimization of the spherical optical area.
For a metric specified by \(A(r)\), \(B(r)\), and \(C(r)\), the decisive
quantity is \(\mathcal{S}=C/A\). A photon can cross a spherical section only
when its squared impact parameter does not exceed \(\mathcal{S}\) there.
Reaching the horizon requires passage through every section on the inward
path, so the global infimum of \(C/A\) fixes the capture threshold.

This formulation is algebraically equivalent to the standard null-geodesic
and effective-potential calculation. It therefore does not change the shadow
prediction when the conventional analysis includes every potential barrier
and the accessibility of each candidate orbit. The contribution of OAMM is
to make that global step explicit through a single observer-accessible
infimum that also accommodates endpoint limits. It then places the candidate
selection rule, finite-distance area ratio, and perturbative extremal-value
formulas within one optical-area prescription.

When the infimum is attained at a smooth interior point, the construction
recovers \(AC'-CA'=0\). A nondegenerate minimum additionally yields the
ordinary exponentially unstable circular null orbit. Degenerate minima
require a higher-order local analysis and need not be exponentially unstable.
The global formulation supplies the necessary selection rule when several
stationary radii exist. One must compare the values of \(C/A\), not merely
list the roots of the stationarity equation. The same formalism gives the
finite-distance law
\(\sin^{2}\alpha_{\rm sh}=\mathcal{A}_{*}/
\mathcal{A}_{\rm opt}(r_{\rm o})\), which remains meaningful when spatial
infinity is unavailable.

The method also separates two aspects of null propagation. The shadow
silhouette depends only on the temporal and angular metric functions,
whereas the radial metric function enters the coordinate-time instability
rate of the critical ray. That rate also depends on the normalization of the
static time coordinate, so comparisons require a normalization convention or
a locally normalized rate. For small metric deformations, the first-order shadow shift is
obtained from the change in \(C/A\) evaluated at the undeformed minimum. The
first-order displacement of the critical radius is not required.

The examples demonstrate complementary features. The synthetic profile
exercises the central global step explicitly by comparing two admissible
exterior minima and selecting the lower optical-area value. The
Reissner--Nordstr\"om and Bardeen cases test exterior admissibility and
reproduce the expected single controlling exterior critical radius for the
representative parameters. The charged dilaton geometry, evaluated in both
nonareal and areal coordinates, confirms that the construction is insensitive
to the radial label and requires the true angular function \(C(r)\). The
Kottler case shows that the photon-sphere radius can remain fixed while the
finite-distance shadow angle changes, and it emphasizes the optical-area
ratio as the appropriate observable in a nonasymptotically flat static
region.

The present analysis is restricted to vacuum propagation in static spherical
spacetimes. Extending the global area principle to frequency-dependent
optical media or to weakly nonspherical geometries is a natural direction for
future work.

\section*{Acknowledgement}
A. \"O. and R. P. would like to acknowledge networking support of the COST Action CA21106 - COSMIC WISPers in the Dark Universe: Theory, astrophysics and experiments (CosmicWISPers), the COST Action CA22113 - Fundamental challenges in theoretical physics (THEORY-CHALLENGES), the COST Action CA21136 - Addressing observational tensions in cosmology with systematics and fundamental physics (CosmoVerse), the COST Action CA23130 - Bridging high and low energies in search of quantum gravity (BridgeQG), and the COST Action CA23115 - Relativistic Quantum Information (RQI) funded by COST (European Cooperation in Science and Technology). R. P. and A. \"O. would also like to acknowledge the funding support of SCOAP3. A. \"O. also thanks to EMU, TUBITAK, ULAKBIM (Turkiye)

\bibliography{ref}

\end{document}